\documentclass [11pt]{article}
\usepackage{amsfonts}
\usepackage{amsmath}
\newtheorem{theorem}{Theorem}[section]

\newtheorem{lemma}{Lemma}

\date{}
\begin{document}
\title{Janossy Densities II. Pfaffian Ensembles.
\footnote{ 
keywords and phrases: random matrices, orthogonal polynomials, 
Janossy densities, pfaffian ensembles
}}
\author{ 
Alexander Soshnikov\thanks{
Department of Mathematics,
University of California at Davis, 
One Shields Ave., Davis, CA 95616, USA.
Email address: soshniko@math.ucdavis.edu.
Research was supported in part by the Sloan Research Fellowship and the 
NSF grant DMS-0103948. 
}  }
\date{}
\maketitle
\begin{abstract}
We extend the main result of the companion paper math-ph/0212063
to the case of the pfaffian 
ensembles.
\end{abstract}

\section{Introduction and Formulation of Results}

Let us consider a $ 2n $-particle pfaffian ensemble introduced by Rains in 
\cite{R}:
Let $ ( X , \lambda) $ be a measure space, $ \phi_1, \phi_2, \ldots \phi_{2n}
 $ 
be complex-valued functions on $ X, $ and $ \epsilon(x,y) $ be an 
antisymmetric kernel
such that 
\begin{equation}
\label{density}
p(x_1, \ldots, x_{2n})= (1/Z_{2n}) \* \det( \phi_j(x_k))_{j,k=1,\ldots 2n} 
\  pf(\epsilon(x_j,x_k))_{j,k=1,\ldots,2n}
\end{equation}
defines the density of a $2n$-dimensional probability distribution on 
$ X^{2n}=X \times \cdots \times X $ with respect to the product measure
$ \lambda^{\otimes 2n}.$  Ensembles of this form were introduced in \cite{R} 
and \cite{TW}.
We recall (see e.g. \cite{G}) that the pfaffian of a $ 2n \times 2n $ 
antisymmetric matrix $ A= (a_{jk}), j,k=1, \ldots, 2n, \ a_{jk}=-a_{kj}, $
is defined as $ pf(A)= \sum_{\tau} (-1)^{ sign(\tau)} a_{i_1j_1} \times
\cdots \times a_{i_n, j_n}, $
where the summation is over all partitions of the set $ \{1, \ldots, 2m \}$
into 
disjoint pairs $ \{i_1, j_1\}, \ldots, \{i_n, j_n\} $ such that 
$ i_k < j_k, \ k=1, \ldots n$, and $ sign(\tau) $ is the sign of the 
permutation
$(i_1, j_1, \ldots, i_n, j_n).$
The normalization constant in (\ref{density})
(usually called the partition function) 
\begin{equation}
\label{partfun}
Z_{2n}= 
\int_{X^{2n}} \* \det( \phi_j(x_k))_{j,k=1,\ldots 2n} 
\  pf(\epsilon(x_j,x_k))_{j,k=1,\ldots,2n}
\end{equation}
can be shown to be equal $ (2n)! \* pf(M), $ where the $ 2n \times 2n $ 
antisymmetric matrix $ M= (M_{jk})_{j,k=1, \ldots, 2n}$ is defined as
\begin{equation}
\label{M}
M_{jk}= \int_{X^2} \phi_j(x) \* \epsilon(x,y) \phi_k(y) \lambda(dx) 
\lambda(dy).
\end{equation}
 For the pfaffian ensemble (\ref{density}) one can explicitly calculate 
k-point correlation 
functions\\
$ \rho_k(x_1, \ldots, x_k): = ((2n)!/(2n-k)!) \* \int_{X^{2n-k}} \*
p(x_1,\ldots, x_k, x_{k+1},\ldots, x_{2n}) \*
 d\lambda(x_{k+1}) \ldots d\lambda(x_{2n}), \ \ k=1, \ldots, 2n $ and show 
that they have the 
pfaffian form (\cite{R})
\begin{equation}
\label{corr}
\rho_k(x_1, \ldots, x_k) =  pf (K(x_i,x_j))_{i,j=1,\ldots k},
\end{equation}
where $ K(x,y) $ is the antisymmetric matrix kernel
\begin{equation}
\label{yadroK}
K(x,y)= \left( \begin{array}{cc} \sum_{1\leq j,k \leq 2n} \phi_j(x) 
\* M_{jk}^{-t} \* \phi_k(y) & 
\sum_{1\leq j,k \leq 2n} \phi_j(x) M_{jk}^{-t} (\epsilon \phi_k)(y) \\
\sum_{1\leq j,k \leq 2n} (\epsilon  \phi_j )(x) M_{jk}^{-t} \phi_k(y) 
& -\epsilon(x,y) + \sum_{1 \leq j,k \leq 2n} 
(\epsilon \phi_j)(x) M_{jk}^{-t} (\epsilon \phi_k)(y)
\end{array} \right),
\end{equation}
provided the matrix $ M $ is invertible (by definition 
$ (\epsilon \phi)(x)= \int_X \epsilon(x,y) \* \phi(y) \lambda(dy) $).
If $ X \subset \Bbb R $ and $\lambda $ is absolutely continuous with respect 
to the Lebesgue measure, then the probabilistic meaning of the k-point 
correlation functions is that of the 
density of probability to find an eigenvalue in each infinitesimal
interval around points $ x_1, x_2, \ldots x_k $. In other words
\begin{eqnarray*}
& & \rho_k(x_1, x_2, \ldots x_k) \lambda(dx_1) \cdots \lambda(dx_k)= \\
& & \Pr \, \{ \textrm{ there is a particle in each infinitesimal interval } \ \ 
(x_i, x_i +dx_i) \}. 
\end{eqnarray*}

On the other hand, if $\mu$ is supported by a discrete set of points, then
$$ \rho_k(x_1, x_2, \ldots x_k) \lambda(x_1) \cdots \lambda(x_k)= 
\Pr \, \{ \textrm{ there is a particle at each
of the points}\ x_i,\ i=1, \ldots, k \}.
$$

In general, random point processes with the k-point correlation functions
of the pfaffian form (\ref{corr}) are called pfaffian random point 
processes. 
(\cite{O}). Pfaffian point processes include determinantal point processes 
(\cite{S})
as a particular case when the matrix kernel has the form 
$ \left( \begin{array}{cc} \epsilon & K \\ -K & 0  \end{array} \right)$
where $K$ is a scalar kernel and $ \epsilon $ is an antisymmetric kernel.

So-called Janossy densities $ \mathcal{J}_{k,I}(x_1, \ldots, x_k), 
\ \ k=0,1,2, \ldots, $
describe the distribution of the eigenvalues in any given interval $I$. If
$ X \subset \Bbb R $ and $ \lambda $ is absolutely continuous with respect to
the Lebesgue measure then  
$$ 
\begin{aligned}
\mathcal{J}_{k,I}(x_1, \ldots x_k) \lambda(dx_1) \cdots \lambda(dx_k)
=\,  \Pr \, & \{\textrm{
there are 
exactly $k$
particles in  $I$,}\\
{} &\textrm{
one in each of the $k$ distinct infinitesimal 
intervals} (x_i, x_i +dx_i) \}
\end{aligned}.
$$
If $ \lambda $ is discrete then
$$ \mathcal{J}_{k,I}(x_1, \ldots x_k) 
=\,  \Pr \{ \textrm{
there are 
exactly $k$
particles in  $I$, one at each of the $k$  
points} \ x_i
, \ i=1, \ldots, k \}.
$$
See \cite{DVJ} and \cite{BoSo} for details and additional discussion.
For pfaffian point processes the Janossy densities also have the
pfaffian
form (see \cite{R}, \cite{O}) with an antisymmetric matrix kernel $ L_I$ :
\begin{equation}
\label{Janossy}
\mathcal{J}_{k,I}(x_1, \ldots, x_k)= const(I) \* 
pf(L_I(x_i,x_j))_{i,j=1,\ldots k},
\end{equation}
where 
\begin{equation}
\label{L}
 L_I= K_I(Id +J\*K_I)^{-1}, 
\end{equation}
$ J= \left( \begin{array}{cc} 0 & 1 \\
-1 & 0 \end{array} \right) $
and 
$const(I)= pf(J - K_I)$ is the Fredholm pfaffian of the 
restriction of the operator 
$K$ on the interval $I$, i.e. 
$const(I) = pf(J-K_I)
= (pf (J +L_I))^{-1}= (\det(Id + J\* K_I))^{1/2}= 
(\det(Id- J \* L_I))^{-1/2}.$ (we refer the reader to \cite{R}, section 8 for 
the treatment of Fredholm pfaffians).

Let us define three $ 2n \times 2n $ matrices $ G^I, \ 
M^I, \ M^{X \setminus I} $:
\begin{equation}
\label{GI}
G^I_{jk} = \int_I \phi_j(x) \int_X  \epsilon(x,y) \* \phi_k(y) \lambda(dy) 
\lambda (dx),
\end{equation}
\begin{eqnarray}
\label{Ms}
M^{I}_{jk}&=&
\int_{I^2} \phi_j(x) \* \epsilon(x,y) \phi_k(y) \lambda(dx) \lambda(dy), \\
M^{X \setminus I}_{jk}&=&
\int_{(X\setminus I)^2} \phi_j(x) \* \epsilon(x,y) \phi_k(y) \lambda(dx) 
\lambda(dy) 
\end{eqnarray}
(please compare (9)-(10) with the above formula (\ref{M}) for $M$).
Throughout the paper we will assume that the matrices $ M^I $ and 
$ M^{X \setminus I} $ are invertible.

The main result of this paper is
\begin{theorem}
The kernel $L_I$ has a form similar to the formula (\ref{yadroK}) for $ K$.
Namely, $ L_I$ is equal to 
\begin{eqnarray}
& & L_I(x,y)= \nonumber \\
\label{yadroL}
& & \left( \begin{array}{cc} \sum_{1\leq j,k \leq 2n} \phi_j(x) 
\* (M^{X \setminus I})^{-t}_{jk} \* \phi_k(y) & 
\sum_{1\leq j,k \leq 2n} \phi_j(x) (M^{X \setminus I})^{-t}_{jk}
(\epsilon_{X \setminus I}  \phi_k)(y) \\
\sum_{1\leq j,k \leq 2n} (\epsilon_{X \setminus I}  \phi_j )(x) 
(M^{X \setminus I})^{-t}_{jk} \phi_k(y) 
& -\epsilon(x,y) + \sum_{1 \leq j,k \leq 2n} 
(\epsilon_{X \setminus I}  \phi_j)(x) (M^{X \setminus I})^{-t}_{jk} 
(\epsilon_{X \setminus I}  \phi_k)(y)
\end{array} \right),
\end{eqnarray}
where $ (\epsilon_{X \setminus I} \* \phi)(x)= \int_{X \setminus I} \epsilon(x,y) \* \phi(y) \lambda(dy).$
\end{theorem}
Comparing (\ref{yadroL}) with (\ref{yadroK}) one can see that the kernel $ L_I$
is constructed in the following way:  1) first it is constructed on 
$ X \setminus I $ by the same recipe used to construct the kernel $ K $ on the 
whole $ X $,  2) it is extended then to $ I $ (we recall that $ L_I $ acts
on $ L^2 (I, d\lambda(x)) $, not on $ L^2 (X \setminus I, d\lambda(x))$).

This result contains as a special case Theorem 1.1 from the companion paper 
\cite{BoSo}.
The rest of the paper is organized as follows. We discuss some interesting 
special cases of the theorem, namely  so-called
polynomial ensembles 
($ \beta=1,2$ and $ 4$) 
in section 2. The proof of the theorem is given in section 3.

\section{Random Matrix Ensembles with $ \beta =1,2,4.$}
We follow the discussion in \cite{R} 
(see also \cite{TW} and \cite{Wid}).

{\bf Biorthogonal Ensembles }.

Consider the particle space to be the union of  two identical measure spaces
$ (V, \mu) $ and $ (W, \mu)$ : 
$ X = V \cup W, \  V=W. $  The configuration of $ 2n $ particles
in $ X $  will consist of $ n $ particles $ v_1, \ldots, v_n $ in $ V $ and 
$ n $ particles $ w_1, \ldots, w_n $ in 
$ W$ in such a way that the configurations of particles in $ V $ and $ W $ 
are 
identical ( i.e. $ v_j = w_j , j=1, \ldots, n $).
Let $ \xi_j, \ \psi_j, j=1, \ldots, n $ be some functions on 
$V$. We define $ \{\phi_j \}$ and $  \epsilon $ in (\ref{density})
so that
$ \phi_j(v)=0, \  v \in V, \ \phi_j(w)= \xi_j(w), \ w \in W, 
\ j=1,\ldots n, \ \phi_j(v)=\psi_{j-n-1}(v), \ v \in V, 
\ \phi_j(w)=0, \ w \in W, \ j=n+1, \ldots, 2n, $ and $
\epsilon(v_1, v_2)=0, \ v_1, v_2 \in V, \ 
\ \epsilon(w_1, w_2)=0, \ w_1, w_2 \in W, \ 
\epsilon(v,w)= - \epsilon(w,v)= \delta_{vw}, \ v \in V, w \in W.
$  The restriction of the measure $ \lambda $ on both $ V $ and $ W $ 
is defined to be equal to $ \mu. $
Then (\ref{density}) specializes into (see Corollary 1.5. in \cite{R})
\begin{equation}
\label{biorthogonal}
p(v_1, \ldots, v_n) = const_n \* 
\det(\xi_j(v_i))_{i,j=1, \ldots, n} \* \det(\psi_j(v_i))_{i,j=1, \ldots, n}.
\end{equation}
Ensembles of the form (\ref{biorthogonal}) are known as biorthogonal ensembles
(see \cite{Mut}, \cite{Bor}).
The statement of the Theorem 1.1
in the case (\ref{biorthogonal}) has been proven 
in the companion paper \cite{BoSo}. The special case of the biorthogonal 
ensemble
(\ref{biorthogonal}) when $ V= \Bbb R$, $ \xi_j(x)=\psi_j(x)= x^{j-1}
$, and $ V = \{\Bbb C\,|\, |z|=1 \}, \xi_j(z) =\overline{\psi}_j(z)=
z^{j-1}$, such ensembles are well known in Random Matrix Theory
as {\it unitary ensembles}, see \cite{M} for details. An ensemble  of the 
form (\ref{biorthogonal}) which is different from random matrix ensembles was
studied in \cite{Mut}. We specifically want to single out the polynomial
ensemble with $ \beta=2.$

{\bf Polynomial $(\beta=2)$  Ensembles}.

Let $ X = \Bbb R $ or $ \Bbb Z, \  
\phi_j(x)= x^{j-1}, \ j=1,\ldots, 2n, $
and $ \lambda(dx) $ has a density $ \omega(x)$
with respect to the reference measure on $ X $ (Lebesgue measure in the 
continuous case, counting measure in the discrete case).
Then the formula (\ref{biorthogonal}) specializes into
\begin{equation}
\label{beta2}
p(v_1, \ldots, v_{n})= const_n \* 
\prod_{1 \leq i < j \leq n}  (v_i - v_j)^2
\* \prod_{1\leq j \leq n} \omega(v_j).
\end{equation}

{\bf Polynomial $(\beta=1)$  Ensembles}.

Let $ X = \Bbb R $ or $ \Bbb Z, \  
\phi_j(x)= x^{j-1}, \ j=1,\ldots, 2n, \ \epsilon(x,y)= 
\frac{1}{2} \* sgn(y-x)$  and $ \lambda(dx) $ has a density $ \omega(x)$
with respect to the reference measure on $ X $ (Lebesgue measure in the 
continuous case, counting measure in the discrete case).
Then the formula
(\ref{density}) specializes into the formula for the density of the joint 
distribution of $ 2n $ particles in a so-called 
$ \beta=1$ polynomial ensemble (see \cite{R},
Remark 1):
\begin{equation}
\label{beta1}
p(x_1, \ldots, x_{2n}) = const_n \* 
\prod_{1 \leq i < j \leq 2n}  |x_i - x_j|
\* \prod_{1\leq j \leq 2n} \omega(x_j).
\end{equation}
In Random Matrix Theory the ensembles (\ref{beta1}) in the continuous case 
are known as {\it orthogonal ensembles} , see \cite{M}.

{\bf Polynomial $(\beta=4)$ Ensembles}.

Similar to the biorthogonal case ($ \beta=2) $ let us
consider the particle space to be the union of  two identical measure spaces
$ (Y, \mu), (Z, \mu), \ \ 
X = Y\cup Z , \ \ Y=Z$, where $ Y = \Bbb R $ or $ Y= \Bbb Z $.
The configuration of $ 2n $ particles $ x_1, \ldots, x_{2n}, $
in $ X $  will consist of $ n $ particles $ y_1, \ldots, y_n $
in $ Y $ and $ n $ particles  $ z_1, \ldots, z_n, $ 
in $ Z$ in such a way that the configurations of particles in $ Y $ and $ Z $ 
are identical.  We define $ \{\phi_j \} $ and $\epsilon $ 
so that
$\phi_j(y) = y^j, \  \in Y, \ 
\phi_j(z)= j\* z^{j-1}, \ z \in Z,\ \  \epsilon(y_1, y_2)=0, 
\ \epsilon(z_1, z_2)=0, \ \epsilon(y,z)= - \epsilon(z,y)= \delta_{yz}. $
As above we assume that the measure $ \mu $ has a density $ \omega $ with 
respect to the reference measure on $ Y.$
Then the formula (\ref{density}) specializes into the formula
for the density of the joint distribution of $ n $ particles
in a $ \beta=4 $ polynomial ensemble (see Corollary 1.3. in \cite{R}))
\begin{equation}
\label{beta4}
p(y_1, \ldots, y_{n}) = const_n \* 
\prod_{1 \leq i < j \leq n}  (y_i - y_j)^4
\* \prod_{1 \leq j \leq n} \omega(y_j).
\end{equation}
In Random Matrix Theory the ensembles (\ref{beta4}) are known as 
{\it symplectic ensembles}, see \cite{M}.

\section{Proof of the Main Result}
Consider matrix kernels 
\begin{equation}
\label{KJLJ}
\mathcal{K}_I=-J \* K_I, \ \ \mathcal{L}_I= -J \*
L_I.
\end{equation}
The the relation (\ref{L}) simplifies into 
\begin{equation}
\label{KL}
\mathcal{L}_I= \mathcal{K}_I \* (Id - \mathcal{K}_I)^{-1}
\end{equation}
which is the same relation that is satisfied by the correlation and Janossy scalar 
kernels
in the determinantal case (\cite{DVJ}, \cite{BO}).
The consideration of $\mathcal{K}_I$ and $\mathcal{L}_I$ is motivated by the fact that
the pfaffians of the $ 2k \times 2k $ matrices with the antisymmetric matrix 
kernels $ K_I$ and
$L_I$ are equal to the quaternion determinants (\cite{M}) of $ 2k \times 2k $
matrices with the kernels
$\mathcal{K}_I, \ \mathcal{L}_I $ when the latter matrices are viewed as 
$ k \times k $ 
quaternion matrices (i.e. each quaternion entry corresponds to  a $ 2\times 2 $ block
with complex entries). It follows from (\ref{yadroK}) and (\ref{KJLJ}) 
that the kernel 
$ \mathcal{K}_I $ is given by the formula
\begin{equation}
\label{kernelK}
\mathcal{K}_I = \sum_{j,k=1,\ldots 2n} M_{jk}^{-t} \left( \begin{array}{cc}
-(\epsilon \phi_j) \otimes \phi_k &  - (\epsilon \phi_j) \otimes
(\epsilon \phi_k) \\ \phi_j \otimes \phi_k & \phi_j \otimes (\epsilon \phi_k)
\end{array} \right) + \left(\begin{array}{cc} 0 & \epsilon \\ 0 & 0 
\end{array} \right).
\end{equation}
Let us denote by $ \widetilde{{\mathcal L}}_I $ the following kernel
\begin{equation}
\label{kernelL}
\widetilde{{\mathcal L}}_I(x,y)= \sum_{1\leq j,k \leq 2n} 
(M^{X \setminus I})^{-t}_{jk} 
\left( \begin{array}{cc} 
-(\epsilon_{X \setminus I}  \phi_j ) \otimes \phi_k 
& -(\epsilon_{X \setminus I}  \phi_j)\otimes \
(\epsilon_{X \setminus I} \phi_k) \\
\phi_j \otimes \phi_k & 
\phi_j \otimes (\epsilon_{X \setminus I}  \phi_k) 
\end{array} \right) + \left( \begin{array}{cc} 0 & \epsilon\\
0 & 0 \end{array} \right).
\end{equation}
As above, 
$ \epsilon \phi $ stands for $ \int_X \epsilon(x,y) \phi(y)$. We use 
the notation  $\phi_j \otimes \phi_k$ is a shorthand for 
$\phi_j(x)\*\phi_k(y).$
To prove the main result of the paper we will show that
$\widetilde{{\mathcal L}}_I = {\mathcal K}_I \* (Id - {\mathcal K}_I)^{-1} $ (in 
other words we are going to prove that $\widetilde{{\mathcal L}}_I = 
{\mathcal L}_I, $ where $ {\mathcal L}_I $ is defined in (\ref{KL}) ).
The proof relies on Lemmas 1 and 2  given below.
Let us introduce the notation
 $ (\epsilon_I \phi)(x)= \int_I \epsilon(x,y) \* \phi_s(y) d\lambda(y). $
We will show that the finite-dimensional subspace $ \mathcal{H}
= Span \left\{ \left(\begin{array}{c} \epsilon \phi_s \\ -\phi_s \end{array}
\right), \ \left(\begin{array}{c} -\epsilon \phi_s \\ -\phi_s \end{array}\right
), \ \left(\begin{array}{c} \epsilon_I \phi_s \\ 0 \end{array}\right) 
\right\}_{s=1,\ldots 2n} $ is invariant under $ \mathcal{K}_I $ and 
$\widetilde{{\mathcal L}}_I.$ The main part of the proof of the theorem is to 
show
that $
\widetilde{{\mathcal L}}_I= \mathcal{K}_I \* (Id - \mathcal{K}_I)^{-1} $ 
holds on 
$\mathcal{H}.$
\begin{lemma}
The operators $\mathcal{K}_I, \ \widetilde{{\mathcal L}}_I$ leave 
$\mathcal{H}$ invariant
and 
$\widetilde{{\mathcal L}}_I= 
\mathcal{K}_I \* (Id - \mathcal{K}_I)^{-1} $ holds on 
$\mathcal{H}$.
\end{lemma}
Below we give the proof of the lemma. 
Using the notations introduced above in (8)-(10)
one can easily calculate
\begin{eqnarray}
\label{odin}
\mathcal{K}_I \* \left( \begin{array}{c} \epsilon \phi_s \\ 0 \end{array}
\right)& = & \sum_{j=1, \ldots 2n} - ((G^I)^{t} \* M^{-1})_{sj} \* 
\left( \begin{array}{c} \epsilon \phi_j \\ -\phi_j \end{array}\right) \\
\mathcal{K}_I \* \left( \begin{array}{c}  0 \\ - \phi_s \end{array}
\right)& = & \sum_{j=1, \ldots 2n} ( G^I \* M^{-1})_{sj} \* 
\left( \begin{array}{c} \epsilon \phi_j \\ -\phi_j \end{array}\right)  - 
\left( \begin{array}{c} \epsilon_I \phi_s \\ 0 \end{array}\right).
\end{eqnarray}
Defining the  $ 2n \times 2n $ matrix $ T $ as
\begin{equation}
\label{T}
T_{sk}= \int_I \phi_s(x) \int_{X \setminus I} \epsilon(x,y) \phi_k(y) 
d\lambda(y) d\lambda(x)
\end{equation}
we compute
\begin{equation}
\label{dva}
\mathcal{K}_I \* \left( \begin{array}{c} \epsilon_I \phi_s \\ 0 \end{array}
\right) =  \sum_{j=1, \ldots 2n} ( (G^I -T)\* M^{-1})_{sj} \* 
\left( \begin{array}{c} \epsilon \phi_j \\ -\phi_j \end{array}\right), 
\end{equation}
where $ (G^I - T)_{sk} = M^I_{sk}= 
\int_{I^2} \phi_s(x) \* \epsilon(x,y) \phi_k(y) \lambda(dx) \lambda(dy).$
One can rewrite the equations (20)-(21) as
\begin{eqnarray}
\label{tri}
\mathcal{K}_I \* \left( \begin{array}{c} \epsilon \phi_s \\ -\phi_s \end{array}
\right)& = & \sum_{j=1, \ldots 2n} ( (G^I -(G^I)^t) \* M^{-1})_{sj}^{t} \* 
\left( \begin{array}{c} \epsilon \phi_j \\ -\phi_j \end{array}\right)
-  \left( \begin{array}{c} \epsilon_I \phi_s \\ 0 \end{array}\right), \\
\mathcal{K}_I \* \left( \begin{array}{c} -\epsilon \phi_s \\ - \phi_s 
\end{array}
\right)& = & \sum_{j=1, \ldots 2n} ( (G^I + (G^I)^t) \* M^{-1})_{sj} \* 
\left( \begin{array}{c} \epsilon \phi_j \\ -\phi_j \end{array}\right)  - 
\left( \begin{array}{c} \epsilon_I \phi_s \\ 0 \end{array}\right)
\end{eqnarray}
We conclude that that the subspace
$ \mathcal{H}$
is indeed invariant
under $ \mathcal{K}_I $ and the matrix of the 
restriction of $\mathcal{K}_I$ on 
$\mathcal{H}$ 
has the following block 
structure
in the basis
$\left \{ \left(\begin{array}{c} \epsilon \phi_s \\ -\phi_s \end{array}
\right), \ \left(\begin{array}{c} -\epsilon \phi_s \\ -\phi_s \end{array}\right
), \ \left(\begin{array}{c} \epsilon_I \phi_s \\ 0 \end{array}\right) \right \}_{s=1,
\ldots 2n} $ : 
\begin{equation}
\label{blockK}
\left( \begin{array}{ccc} (G^I-(G^I)^t)\* M^{-1} &(G^I+(G^I)^t)\* M^{-1} & 
                          (G^I-T)\* M^{-1} \\ 0 & 0 & 0 \\ -Id & -Id & 0
\end{array} \right)
\end{equation}
(in particular $ Ran (\mathcal{K}_I |_{\mathcal{H}})= Span 
\left \{ \left(\begin{array}{c} \epsilon \phi_s \\ -\phi_s \end{array}
\right), \  \left(\begin{array}{c} \epsilon_I \phi_s \\ 0 
\end{array}\right) \right \}_{s=1,
\ldots 2n} $  ).
Let us introduce some additional notations:
\begin{eqnarray}
\label{ABC}
 A &=& (G^I-(G^I)^t)\* M^{-1}, \\ 
 B &=& (G^I+(G^I)^t)\* M^{-1}, \\
 C &=& (G^I-T)\* M^{-1}.
\end{eqnarray}
When
a matrix has a block form 
$
\mathcal{M}= \left( \begin{array}{ccc}  A   & B & C \\
                           0   & 0 & 0 \\
                           -Id   & -Id &  0 
\end{array} \right)
$
(as it is in our case)
the matrix $ \mathcal{M} \* (Id- \mathcal{M})^{-1} $
has the block form
\begin{equation}
\label{resolvent}
\left( \begin{array}{ccc}  (Id-A+C)^{-1} -Id   & (B-C)(Id-A+C)^{-1}  & 
                           C(Id-A+C)^{-1} \\ 0 & 0 & 0\\
                            -(Id-A+C)^{-1} & -Id-(B-C)(Id-A+C)^{-1} &
                             -C (Id-A+C)^{-1} 
\end{array} \right)
\end{equation}
As one can see from the formulas (31)-(33) the invertibility
of $ Id - \mathcal{M} $ follows from the invertibility of $ M^{X\setminus I}$
which has been assumed throughout the paper.
We have 
\begin{eqnarray}
\label{uravneniya}
(Id-A+C)^{-1} &=& M \* ( M + (G^I)^t -T)^{-1}= M \* (M^{X \setminus I})^{-1}
 \\
C \* (Id -A +C)^{-1} &=& (G^I -T) \* ( M + (G^I)^t -T )^{-1} 
= M^I \* (M^{X \setminus I})^{-1} \\
(B-C) \* (Id -A +C)^{-1} &=& ( (G^I)^t +T) (M + (G^I)^t -T)^{-1}=
((G^I)^t +T) \* (M^{X \setminus I})^{-1}.
\end{eqnarray}
Let us now compute the matrix of the restriction of 
$ \widetilde{{\mathcal L}}_I $ on
$\mathcal{H}.$  We have
\begin{equation}
\label{kernelL}
\widetilde{{\mathcal L}}_I  
= \sum_{j,k=1,\ldots 2n} (M^{X \setminus I}_{jk})^t \left( \begin{array}{cc}
-(\epsilon_{X \setminus I} \phi_j) \otimes \phi_k & 
 - (\epsilon_{X \setminus I} \phi_j) \otimes
(\epsilon_{X \setminus I} \phi_k) \\ \phi_k \otimes \phi_k & \phi_j 
\otimes (\epsilon_{X \setminus I} \phi_k)
\end{array} \right) + \left(\begin{array}{cc} 0 & \epsilon_{X \setminus I} \\
                                              0 & 0
\end{array} \right).
\end{equation}
Similarly to the computations above one can see that $ \mathcal{H} $ is 
invariant under $\widetilde{{\mathcal L}}_I$ and
\begin{eqnarray}
\label{chetyre}
\widetilde{{\mathcal L}}_I \* \left( \begin{array}{c} \epsilon \phi_s \\ -\phi_s \end{array}
\right)& = & \sum_{j=1, \ldots 2n} ( (T -(G^I)^t) \* 
(M^{X\setminus I})^{-1})_{sj} \* 
\left( \begin{array}{c} \epsilon \phi_j \\ -\phi_j \end{array}\right) 
\nonumber \\
 & & -  \sum_{1\leq j \leq 2n} ((T -(G^I)^t) \* 
(M^{X\setminus I})^{-1})_{sj} \* 
\left( \begin{array}{c} \epsilon_I \phi_j \\ 0 \end{array}\right)
-  \left( \begin{array}{c} \epsilon_I \phi_s \\ 0 \end{array}\right),   \\
\widetilde{{\mathcal L}}_I \* \left( \begin{array}{c} -
\epsilon \phi_s \\ - \phi_s 
\end{array}
\right) 
& = & \sum_{j=1, \ldots 2n} ( (T + (G^I)^t) \* 
(M^{X \setminus I})^{-1})_{sj} \* 
\left( \begin{array}{c} \epsilon \phi_j \\ -\phi_j \end{array}\right) 
\nonumber \\
 & & -  \sum_{1\leq j \leq 2n} ((T +(G^I)^t) \* 
(M^{X\setminus I})^{-1})_{sj} \* 
\left( \begin{array}{c} \epsilon_I \phi_j \\ 0 \end{array}\right)
-  \left( \begin{array}{c} \epsilon_I \phi_s \\ 0 \end{array}\right), 
\end{eqnarray}
and
\begin{equation}
\label{pyat}
\widetilde{{\mathcal L}}_I \* \left( \begin{array}{c} \epsilon_I \phi_s \\ 0 \end{array}
\right) =  \sum_{j=1, \ldots 2n} ( (G^I -T)\* 
(M^{X \setminus I})^{-1})_{sj} \* 
\left( \begin{array}{c} \epsilon \phi_j \\ -\phi_j \end{array}\right) 
-\sum_{j=1, \ldots 2n} ( (G^I -T)\* 
(M^{X \setminus I})^{-1})_{sj} \* 
\left( \begin{array}{c} \epsilon_I \phi_j \\ 0 \end{array}\right).
\end{equation}
Therefore the restriction of $ \widetilde{{\mathcal L}}_I$ to $\mathcal{H}$ 
in the basis
$\left \{ \left(\begin{array}{c} \epsilon \phi_s \\ -\phi_s \end{array}
\right), \ \left(\begin{array}{c} -\epsilon \phi_s \\ -\phi_s \end{array}\right
), \ \left(\begin{array}{c} \epsilon_I \phi_s \\ 0 \end{array}\right) \right \}_{s=1,
\ldots 2n} $  
has the following block structure
\begin{equation}
\label{blockL}
\left( \begin{array}{ccc} (T-(G^I)^t)\* (M^{X \setminus I})^{-1} & 
                          (T+(G^I)^t)\* (M^{X\setminus I})^{-1} & 
                          (G^I-T)\* (M^{X \setminus I})^{-1} \\
                           0 & 0 & 0 \\
-Id -(T-(G^I)^t)\* (M^{X \setminus I})^{-1}  &
-Id -(T+(G^I)^t)\* (M^{X\setminus I})^{-1} &
-(G^I-T)\* (M^{X \setminus I})^{-1} 
\end{array} \right)
\end{equation} 
Comparing (\ref{resolvent}), (31)-(33) and (\ref{blockL})
we see that $ \widetilde{{\mathcal L}}_I = 
\mathcal{K}_I \* (Id - \mathcal{K}_I)^{-1}$
on $\mathcal{H}$. Lemma is proven.

To show that $ \widetilde{{\mathcal L}}_I =\mathcal{K}_I \* 
(Id - \mathcal{K}_I)^{-1}$  also holds on the 
complement of 
$ \mathcal{H} $ it is enough to prove  it  on the subspaces 
$ \left( \begin{array}{c}
(\mathcal{H}_1)^{\bot} \\ 0 \end{array} \right),$ and
$ \left( \begin{array}{c} 0\\
(\mathcal{H}_2)^{\bot} \end{array} \right) $,
where  $ \mathcal{H}_1 = 
Span(\overline{\epsilon_I \phi_s})_{k=1, \ldots, 2n} $ and $ \mathcal{H}_2 =
Span(\overline{\phi_s})_{k=1, \ldots, 2n}$. The 
inveribility of the matrix
$ M_I$ implies  that actually it is enough to prove
$ \widetilde{{\mathcal L}}_I =\mathcal{K}_I \* 
(Id - \mathcal{K}_I)^{-1}$  on the subspaces
$ \left( \begin{array}{c}
(\mathcal{H}_2)^{\bot} \\ 0 \end{array} \right),$ and
$ \left( \begin{array}{c} 0\\
(\mathcal{H}_1)^{\bot} \end{array} \right) $.
Here we use the standard notation $ (\mathcal{H}_i)^{\bot} $ 
for the orthogonal 
complement in $ L^2(I)$ with the standard scalar product $ (f,g)_I =
\int_I \* \overline{f(x)} \* g(x) \* d\lambda(x).$ We start with the first 
subspace.
\begin{lemma}
The relation $ \widetilde{{\mathcal L}}_I = 
\mathcal{K}_I \* (Id - \mathcal{K}_I)^{-1} $ 
holds on $ \left( \begin{array}{c} 0 \\ (\mathcal{H}_1)^{\bot} 
\end{array} \right). $
\end{lemma}
The proof is a straightforward check. The notations are slightly simplified 
when the functions $ \{ \epsilon_I \phi_k , \ \epsilon \phi_k, \ k=1,\ldots,
2n \}$
are linearly independent  in $L^2(I)$. The degenerate case is left to the 
reader.
Consider $ f_s \in (\mathcal{H}_1)^{\bot},  \ s=1, \ldots , 2n $ such that
\begin{equation}
\label{scalarprod}
(\overline{\epsilon \phi_k}, f_s)_I = (\overline{\epsilon \phi_k}, 
\phi_s)_I, \ k=1, \ldots, 2n.
\end{equation}
We are going to
establish the relation for
$ \left(\begin{array}{c}
0 \\ f_s  \end{array} \right), $ which then immediately extends by linearity 
to the linear combinations of $ \left(\begin{array}{c}
0 \\ f_s  \end{array} \right).$
We write
\begin{eqnarray}
\label{o}
\mathcal{K}_I \* \left( \begin{array}{c} 0 \\ -f_s \end{array}
\right)& = & \sum_{j,k=1, \ldots 2n}  M^{-t}_{jk}\* (\overline{\epsilon 
\phi_k}, -f_s)_I
\* 
\left( \begin{array}{c} -\epsilon \phi_j \\ \phi_j \end{array}\right) -
\left( \begin{array}{c} \epsilon_I f_s \\ 0 \end{array}\right) \nonumber \\
& = & \sum_{j,k=1, \ldots 2n}  M^{-t}_{jk}\* (\overline{\epsilon \phi_k}, 
-\phi_s)_I
\*
\left( \begin{array}{c} -\epsilon \phi_j \\ \phi_j \end{array}\right) -
\left( \begin{array}{c} \epsilon_I f_s \\ 0 \end{array}\right) \nonumber \\
& = & \sum_{j=1, \ldots 2n} (G^I \* M^{-1}_{sj})
\*
\left( \begin{array}{c} \epsilon \phi_j \\ -\phi_j \end{array}\right) -
\left( \begin{array}{c} \epsilon_I f_s \\ 0 \end{array}\right) 
\end{eqnarray}
(we have used (\ref{scalarprod}) in the second equality)
and
\begin{equation}
\label{oo}
\mathcal{K}_I \* \left( \begin{array}{c} \epsilon_I \phi_s \\ - f_s 
\end{array}
\right) =  \sum_{j=1, \ldots 2n} ( (G^I-(G^I)^t)\* M^{-1})_{sj} \* 
\left( \begin{array}{c} \epsilon \phi_j \\ -\phi_j \end{array}\right)  - 
\left( \begin{array}{c} \epsilon_I f_s \\ 0 \end{array}\right).
\end{equation}
Combining (\ref{o}) and (\ref{oo}) we get
\begin{eqnarray}
\label{ooo}
\mathcal{K}_I \* \left( \begin{array}{c} -\epsilon_I \phi_s \\ -f_s 
\end{array}
\right)& = & \sum_{j=1, \ldots 2n} ( (G^I+(G^I)^t)\* M^{-1})_{sj} \* 
\left( \begin{array}{c} \epsilon \phi_j \\ -\phi_j \end{array}\right)  - 
\left( \begin{array}{c} \epsilon_I f_s \\ 0 \end{array}\right),
\end{eqnarray}
Similarly to (\ref{dva}) we compute
\begin{equation}
\label{oooo}
\mathcal{K}_I \* \left( \begin{array}{c} \epsilon_I \phi_s \\ 0 \end{array}
\right) =  \sum_{j=1, \ldots 2n} ( (G^I -T)\* M^{-1})_{sj} \* 
\left( \begin{array}{c} \epsilon \phi_j \\ -\phi_j \end{array}\right) 
\end{equation}
It should be noted that $ \mathcal{K}_I \* \left( \begin{array}{c} 
\epsilon_I f_s\\ 0 \end{array} \right) =0 $ because 
$ \int_I \*(\epsilon_I \phi_s)(x) \phi_j(x) \* d\lambda(x) =
- \int_I \* f_s(x) \* (\epsilon_I \phi_j)(x) \* d\lambda(x) =0 $
for all
$ j=1,\ldots, 2n.$
This together with (\ref{scalarprod}) allows us to conclude that the 
calculation of 
$ \mathcal{K}_I (Id - \mathcal{K}_I)^{-1}\* \left( \begin{array}{c} 
0 \\  f_s \end{array} \right)  $  is almost identical to the 
calculation of
$ \mathcal{K}_I (Id - \mathcal{K}_I)^{-1}\* \left( \begin{array}{c} 
0 \\ \phi_s \end{array} \right)  $   with the only difference that in the 
former one we have to replace the term
$ - \left( \begin{array}{c} 
\epsilon_I \phi_s\\ 0 \end{array} \right)  $ 
by
$ - \left( \begin{array}{c} 
\epsilon_I f_s\\ 0 \end{array} \right)  $ 
(see the last equation of (\ref{i})). Namely
\begin{eqnarray}
\label{i}
\mathcal{K}_I \* (Id - \mathcal{K}_I)^{-1} \* \left( \begin{array}{c}
0 \\ -f_s \end{array} \right) &=&
\mathcal{K}_I \* (Id - \mathcal{K}_I)^{-1} \* \left( \frac{1}{2} \*
\left( \begin{array}{c} \epsilon \phi_s \\ -f_s \end{array} \right) +
\left( \begin{array}{c} -\epsilon \phi_s \\ -f_s \end{array} \right)
\right) \nonumber \\
&=&  \sum_{j=1, \ldots, 2n} (1/2) \left( (A+B)\*(Id-A+C)^{-1} \right)_{sj} \* 
\left( \begin{array}{c} \epsilon_I \phi_j \\ -\phi_j \end{array} \right)
\nonumber \\
&-& \sum_{j=1, \ldots, 2n} (1/2) \left( (A+B)\*(Id-A+C)^{-1} \right)_{sj} \* 
\left( \begin{array}{c} \epsilon_I \phi_j \\ 0 \end{array} \right) -
\left( \begin{array}{c} \epsilon_I f_s \\ 0 \end{array}\right)
\nonumber \\
&=& \sum_{j=1, \ldots, 2n} 
\left[ G^I \*  (M^{X\setminus I})^{-1} \right]_{sj} \*
\left[ 
\left( \begin{array}{c} \epsilon \phi_j \\ -\phi_j \end{array} \right) 
- \left( \begin{array}{c} \epsilon_I \phi_j \\ 0 \end{array} \right) \right]
- \left( \begin{array}{c} \epsilon_I f_s \\ 0 \end{array}\right).
\end{eqnarray}
where $ A, \ B, \ C $ are defined in (27)-(29).
At the same time
\begin{eqnarray}
\label{iii}
\widetilde{{\mathcal L}}_I\* \left( \begin{array}{c}
0 \\ -f_s \end{array} \right) &=&
\sum_{j,k=1, \ldots,2n} (M^{X\setminus I})^{-t})_{jk} \*
\left( \begin{array}{c} \epsilon_{X \setminus I} \phi_j 
\\ -\phi_j \end{array} \right) \*(\overline{\epsilon_{X \setminus I} \phi_k}
, f_s)_I
- \left( \begin{array}{c} \epsilon_I f_s \\ 0 \end{array}\right) \nonumber \\
&=&
\sum_{j,k=1, \ldots,2n} (M^{X\setminus I})^{-t})_{jk} \*
\left( \begin{array}{c} \epsilon_{X \setminus I} \phi_j 
\\ -\phi_j \end{array} \right) \*(\overline{\epsilon \phi_k}, f_s)_I
- \left( \begin{array}{c} \epsilon_I f_s \\ 0 \end{array}\right) \nonumber \\
&=& \left[ G^I \*  (M^{X\setminus I})^{-1} \right]_{sj} \*
\left[ 
\left( \begin{array}{c} \epsilon \phi_j \\ -\phi_j \end{array} \right) 
- \left( \begin{array}{c} \epsilon_I \phi_j \\ 0 \end{array} \right) \right]
- \left( \begin{array}{c} \epsilon_I f_s \\ 0 \end{array}\right).
\end{eqnarray}
Therefore
$
\widetilde{{\mathcal L}}_I \* \left( \begin{array}{c}
0 \\ -f_s \end{array} \right) = \mathcal{K}_I \* (Id -\mathcal{K}_I)^{-1}
\* \left( \begin{array}{c} 0 \\ -f_s \end{array} \right),
\ \ s=1, \ldots 2n.$
By linearity result follows for all
$ \left(\begin{array}{c}
0 \\ f  \end{array} \right) $ such that 
$ (\overline{\epsilon_I\phi_k},f)_I=\int_I(\epsilon_I
\phi_k)(x)\* f(x) \* d\lambda(x)=0, 
\ \ k,j=1, \ldots 2n.$ Lemma 2 is proven.

To check (\ref{KL}) on $ \left( \begin{array}{c}
(\mathcal{H}_2)^{\bot} \\ 0 \end{array} \right) $ we note
that
$ \mathcal{K}_I \* (Id -\mathcal{K}_I)^{-1}  \* \left( \begin{array}{c}
g \\ 0 \end{array} \right) = \widetilde{{\mathcal L}}_I 
\* \left( \begin{array}{c}
g \\ 0 \end{array} \right) =0 $
for $g $ such that $ \int_I \* g(x) \* \phi_k(x) \* d\lambda(x)  = 0 , 
\ \ k=1, \ldots, 2n, $
which together with the invertibility of $ M $ finishes the proof.
Theorem is proven.

\vspace{0.25cm}
\noindent

{\bf Acknowledgements.}\, 
It is a great pleasure to thank Alexei Borodin and Eric Rains for useful 
discussions. The author would like to thank the anonimous referee for pointing
out to a number of misprints.

\def\cmp{{\it Commun. Math. Phys.} }

\end{document}